\documentclass[prl,twocolumn,showpacs,floatfix]{revtex4-1}

\usepackage{natbib}
\usepackage{graphicx}
\usepackage{dcolumn}
\usepackage{amsmath}
\usepackage{amssymb}
\usepackage{bm}

\def\beq{\begin{equation}}
\def\eeq{\end{equation}}

\def\om{\omega}

\def\eps{\epsilon}

\def\s{\sigma}
\def\D{\Delta}

\def\ad{a^\dagger}

\def\rd{{\rm{d}}}

\def\p{\phi}

\def\bph{\bar{\phi}}

\def\ra{\rightarrow}

\def\Cc{\mathbb{C}}
\def\Zz{\mathbb{Z}}
\def\R{\mathbb{R}}

\def\dd{\frac{\rd}{\rd z}}

\def\Hp{{\cal H}_+}

\def\Zpl{\rm{I\!N}}

\def\A{\bm{A}}

\begin{document}
\title{Note on the Analytical Solution of the Rabi Model}

\author{ D.~Braak}

\affiliation{EP VI and Center for Electronic Correlations and Magnetism, 
University of Augsburg, 86135 Augsburg, Germany}

\date{November 23, 2012}

\begin{abstract}
It is shown that a recent critique (arXiv:1210.1130 and
arXiv:1211.4639) concerning the analytical solution of the Rabi model
is unfounded.
\end{abstract}


\maketitle

It was demonstrated in \cite{db} that
the spectrum of the quantum Rabi Hamiltonian ($\om=\hbar=1, g>0$),
\beq
H_R= \ad a +g\s_x(a+\ad) +\D\s_z. 
\label{ham1}
\eeq
consists of two parts, the regular spectrum with
energy eigenvalues $E^\pm_n+g^2\notin \Zpl$ and
the exceptional spectrum (Juddian solutions) with
$E^{\rm exc}_n+g^2\in \Zpl$, which may occur for
special values of the model parameters $g$ and $\D$. 
The index $\pm$ 
denotes the parity of the regular eigenstate
belonging to $E^\pm_n$. The exceptional states 
(if they are present) are all
doubly degenerate with respect to parity.
The parity invariance of the model ($H_R$ commutes with
$e^{i\pi\ad a}\s_z$) is instrumental to derive the
function
$G_\pm(x)$, 
\beq
G_\pm(x)=\sum_{n=0}^\infty K_n(x)
\left(1\mp\frac{\D}{x-n}\right)g^n,
\label{sol}
\eeq  
where the $K_n(x)$ are known functions of $g,\D$ and $x$.\\
$G_+(x)$ ($G_-(x)$) determines
the regular spectrum of (\ref{ham1}) in the
subspace with positive (negative) parity, because
$G_\pm(x^\pm_n)=0$ for real $x^\pm_n$ if and only if $E^\pm_n+g^2=x^\pm_n$.

Maciejewski {\it et al.} argue in \cite{mac} that this result is invalid because
some of the real zeroes of $G_\pm(x)$ may not correspond to
eigenvalues of (\ref{ham1}). The authors of \cite{mac}
do not dispute the fact that all points of the regular spectrum
correspond to zeroes of (\ref{sol}), but suspect that not all
those zeroes are physical. I shall now prove that this is not the case.

It is sufficient to confine the discussion 
to fixed (positive) parity (negative parity is obtained by
replacing $\D$ with $-\D$ in the subsequent formulas).
The subspace $\Hp$ with positive parity is isomorphic to $\cal B$,
the Bargmann space of analytic functions
(see \cite{db}) and 
the derivation of (\ref{sol}) starts with the following system of
coupled differential equations for the wave function $\psi(z)$,
which solves the Schr\"odinger equation $H_+\psi(z)=E\psi(z)$
in $\Hp$,  
\begin{subequations}
\begin{align}
(z+g)\dd\p_1(z) +(gz-E)\p_1(z) +\D\p_2(z) &= 0\\
(z-g)\dd\p_2(z) -(gz+E)\p_2(z) +\D\p_1(z) &= 0. 
\label{coup-sys}
\end{align}
\end{subequations}
Here we have used the  notation
$\psi(z)=\p_1(z)$, $\psi(-z)=\p_2(z)$. This system corresponds to a
linear homogeneous differential equation of the first order for the vector-valued function
$\Psi(z)=(\p_1(z),\p_2(z))^T$,
\begin{equation}
\dd\left(\!
\begin{array}{c}
\p_1(z)\\
\p_2(z)
\end{array}
\!\right)
=
\left(\!
\begin{array}{cc}
\frac{E-gz}{z+g} & \frac{-\D}{z+g}\\
\frac{-\D}{z-g} &\frac{E+gz}{z-g}
\end{array}
\!\right)
\left(\!
\begin{array}{c}
\p_1(z)\\
\p_2(z)
\end{array}
\!
\right)
\label{sys-1}
\end{equation}
(\ref{sys-1}) has regular singular points at $z=\pm g$ and an irregular
singular point (of rank 1) at infinity.
Now it follows from the symmetry of this equation under the reflection
$z\ra -z$
that the function $\Phi(z)=(\p_2(-z),\p_1(-z))^T$ satisfies
(\ref{sys-1}) as well. $\p_{1,2}(z)$ have power series expansions
around $z=-g$,
\begin{subequations}
\begin{align}
\p_1(z) &= e^{-gz}\sum_{n=0}^\infty K_n(x)\D\frac{(z+g)^n}{x-n},
\label{phi12a}\\
\p_2(z)&= e^{-gz}\sum_{n=0}^\infty K_n(x)(z+g)^n, 
\label{phi12b}
\end{align}
\end{subequations}
where $x$ denotes the spectral parameter $x=E+g^2$.
It follows that $\Psi(z)$ is analytic in an open disk $D_1$ with radius
$2g$ centered at $z=-g$. Likewise, $\Phi(z)$ is analytic in a disk
$D_2$ with the same radius centered at $z=g$.  
All points in $D_0=D_1\cap D_2$ are ordinary points of (\ref{sys-1}) \cite{ince1}.
It means that if $\Psi(z_0)=\Phi(z_0)$ for any $z_0\in D_0$, $\Psi(z)$
and $\Phi(z)$ coincide for all $z\in D_0$. But this  entails
that $\phi_1(z)=\psi(z)$ is analytic in the whole complex plane because
then $\phi_2(-z)$ is its analytic continuation beyond the radius of
convergence of (\ref{phi12a}), comprising the second regular singular
point of (\ref{sys-1}) at $z=g$. It follows that the conditions
\begin{subequations}
\begin{align}
\p_1(z_0) &= \p_2(-z_0)
\label{cond-1a}\\
\p_2(z_0) &= \p_1(-z_0)
\label{cond-1b}
\end{align}
\end{subequations}
for any $z_0\in D_0$ are necessary
{\it and} sufficient for $\psi(z)$ to be an element of the
Bargmann space; the spectral parameter $x$, determined by (\ref{cond-1a}, \ref{cond-1b}),
corresponds therefore to an energy eigenvalue.
Now (\ref{cond-1a}) is equivalent to (\ref{cond-1b}) if $z_0=0 \in D_0$,
from which the expression for $G_\pm(x)$ given in Eq.~(\ref{sol}) 
follows immediately. This completes the proof
sketched in \cite{supp}. In a comment \cite{mac2} to the first version 
of this note, Maciejewski {\it et al.} still 
doubt the validity of the proof by invoking a standard theorem of 
complex analysis which says that a function holomorphic in a bounded, connected
region $D$ of $\Cc$ vanishes identically if it vanishes at a
 denumerable infinity of points within $D$. This theorem has nothing to do 
with the present problem. Here we use the following elementary result \cite{cod} from the theory of linear differential 
equations:\par
Theorem:\ Let the vector-valued function $\bm{f}(z)$ satisfy a linear 
homogeneous differential equation of the first order which has only ordinary points in the connected complex 
domain $D$. If $\bm{f}(z)$ vanishes at some point $z_0\in D$, 
it vanishes everywhere in $D$.\par 
The condition $\bm{f}(z_0)=\Phi(z_0)-\Psi(z_0)=\bm{0}$ corresponds to
(\ref{cond-1a}, \ref{cond-1b}), and both are equivalent for $z_0=0\in D_0$. If 
(\ref{cond-1a}) is satisfied at $z_0=0$, (\ref{cond-1b}) is satisfied as well
and the {\it vector} $\bm{f}(0)$ vanishes. This is enough to conclude
that $\Phi(z)=\Psi(z)$ throughout $D_0$. It is not necessary that one of the {\it components}
of $\bm{f}(z)$ vanishes at two distinct points (see below), but both components
must vanish at {\it one} point. This is equivalent to the
condition $f(z_0)=f'(z_0)=0$ if the scalar $f(z)$ satisfies a second order 
differential equation, as the components of $\bm{f}(z)$ do.  

For $z_0\neq 0$, Eqs. (\ref{cond-1a}) and (\ref{cond-1b}) are not
equivalent and it becomes possible to have 
a solution to (\ref{cond-1a}), while (\ref{cond-1b}) is not satisfied.
This was discovered numerically in \cite{mac} for real $z_0$. 
Clearly, no unphysical solutions were obtained for $z_0=0$, but
this is not a ``lucky'' accident as the
authors of \cite{mac} believe, who checked the zeroes of $G_\pm(x)$
for $x$ up to 30.

The same argument applies to the generalized Rabi model with broken
$\Zz_2$-symmetry. Its Hamiltonian reads,
\beq
H_\eps =  \ad a +g\s_x(a+\ad)+\eps\s_x +\D\s_z.
\label{hamz2}
\eeq
As was shown in \cite{supp}, the eigenvalue equation for
$H_\eps$ is equivalent via integrable embedding to the following
differential equation for the vector-valued function
$\Psi(z)=(\p_1(z),\p_2(z),\bph_1(z),\bph_2(z))^T$,
\beq
\dd\Psi(z)=\A(z)\Psi(z),
\label{sys-2}
\eeq
with the coefficient matrix,
\beq
\A(z)=\left(\!
\begin{array}{cccc}
\frac{E-\eps-gz}{z+g} & 0 & 0 & \frac{-\D}{z+g}\\
0 & \frac{E+\eps-gz}{z+g}&\frac{-\D}{z+g} & 0\\
0 & \frac{-\D}{z-g}&\frac{E-\eps+gz}{z-g} & 0\\
\frac{-\D}{z-g} & 0 & 0 & \frac{E+\eps+gz}{z-g}
\end{array}
\!\right).
\eeq
(\ref{sys-2}) has the same singularity structure as 
(\ref{sys-1}) and regions $D_0,D_1,D_2$ can be defined
as in the symmetric case.
Due to the embedding, Eq.~(\ref{sys-2}) has again a
$\Zz_2$-symmetry, which entails that with $\Psi(z)$ also
the function $\Phi(z)=(\bph_1(-z),\bph_2(-z),\p_1(-z),\p_2(-z))^T$
satisfies (\ref{sys-2}).
After expansion of $\Psi(z)$ in powers of $z$ around the
regular singular point $-g$, the condition $\Psi(z_0)=\Phi(z_0)$ 
for $z_0\in D_0$ leads to the following
set of equations,
\begin{subequations}
\begin{align}
e^{-gz_0}\sum_{n=0}^\infty \frac{\D K_n^-}{x-\eps-n}(z_0+g)^n
=ce^{gz_0}\sum_{n=0}^\infty K_n^+(g-z_0)^n,
\label{ext-1}\\
ce^{-gz_0}\sum_{n=0}^\infty \frac{\D K_n^+}{x+\eps-n}(z_0+g)^n
=e^{gz_0}\sum_{n=0}^\infty K_n^-(g-z_0)^n,
\label{ext-2}\\
ce^{-gz_0}\sum_{n=0}^\infty K_n^+(z_0+g)^n
=e^{gz_0}\sum_{n=0}^\infty \frac{\D K_n^-}{x-\eps-n}(g-z_0)^n,
\label{ext-3}\\
e^{-gz_0}\sum_{n=0}^\infty K_n^-(z_0+g)^n
=ce^{gz_0}\sum_{n=0}^\infty \frac{\D K_n^+}{x+\eps-n}(g-z_0)^n,
\label{ext-4}
\end{align}
\end{subequations}
with an unknown constant $c$. 
The $K_n^\pm$ are known functions of $g,\D$ and $x=E+g^2$. 
For $z_0=0$ it is obvious that
(\ref{ext-1}) is equivalent to (\ref{ext-3}) and
(\ref{ext-2}) to (\ref{ext-4}). We are left with the 
two equations,
\begin{subequations}
\begin{align}
\sum_{n=0}^\infty \left[cK_n^+-\frac{\D}{x-\eps-n}K_n^-\right]g^n&=0\label{ex0-1}\\
\sum_{n=0}^\infty \left[K_n^--\frac{c\D}{x+\eps-n}K_n^+\right]g^n&=0.
\label{ex0-2}
\end{align}
\end{subequations}
Eliminating $c$ from Eqs.~(11), we obtain the $G$-function for the generalized
Rabi model \cite{db}, 
\beq
G_\eps(x)=\D^2{\bar{R}}^+(x){\bar{R}}^-(x)-R^+(x)R^-(x)
\eeq
with
\begin{subequations}
\begin{align}
R^\pm(x)=&\sum_{n=0}^\infty K^{\pm}_n(x)g^n
\label{r-1}\\
{\bar{R}}^\pm(x)=&
\sum_{n=0}^\infty \frac{K^{\pm}_n(x)}{x-n\pm\eps}g^n.
\label{r-2}
\end{align}
\end{subequations}
The reflection symmetry of the extended model allows to reduce the number
of conditions as in the manifestly symmetric case. Therefore,  
the function $W(x,g,\D,\eps)$ derived in \cite{mac} has exactly the same
real zeroes as $G_\eps(x)$ and yields the same spectrum as
seen in Fig.~5 of \cite{mac}. However, the proposed method is
an interesting generalization of the approach introduced in \cite{db}
which could be applicable to cases where embedding into a
symmetric model is not possible. 
 
Regarding the numerical computation
of the spectrum of the quantum Rabi model (\ref{ham1}),
it may be advantageous to define a generalized
$G$-function $G_\pm(x;z)$ by
\beq
G_\pm(x;z)=\p_2(-z)-\p_1(z).
\label{genG}
\eeq 
The vanishing of $G_\pm(x;z_0)$ for $z_0\in D_0$ corresponds
to (\ref{cond-1a}). Interestingly, this condition is sufficient
to determine the spectrum if $\Im(z_0)\neq 0$. To see this,
we note that the conditions (6) correspond to a
two-point boundary value problem for $G_\pm(x;z)$ in the complex plane, namely
$G_\pm(x;z_0)=G_\pm(x;-z_0)=0$. Because $G_\pm(x;z)$ satisfies a linear homogeneous
differential equation of the second order, which is obtained from
Eq.~(\ref{sys-1}) by eliminating $\p_2(z)$, this boundary value problem
is incompatible \cite{ince2} and has only the solution $G_\pm(x;z)\equiv 0$.     
Let $G^j_\pm(x;z)$ for $j=1,2$ denote two linearly independent
solutions of the differential equation satisfied by $G_\pm(x;z)$ 
(Eq.~(11) in \cite{mac}). Let us assume that
$G_\pm(x;z_0)=0$ for $z_0\in D_0$. Clearly,
$G_\pm(x;z_0)^\ast=0$ as well. But because the coefficients of the power series
of $G_\pm(x;z)$ in $z$ are real (see Eq.~(5)), we have
$G_\pm(x;z_0^\ast)=0$, i.e. $G_\pm(x;z)$ vanishes at $z_0$ and $z_0^\ast$.
This is again an incompatible two-point boundary value problem
if $\Im(z_0)\neq 0$, because
$G^1_\pm(x;z_0)G^2_\pm(x;z_0^\ast)\neq G^1_\pm(x;z_0^\ast)G^2_\pm(x;z_0)$ 
for almost all $z_0\in D_0$. It does not preclude isolated points $z_0\in D_0$ which yield
non-trivial solutions to $G_\pm(x;z_0)=G_\pm(x;z_0^\ast)=0$.
In \cite{mac2} the following statement is made:
``The key observation is that the function in question satisfies a second order linear homogeneous
equation so that we only need to make it equal to zero at two {\bf distinct} points.'' This is obviously incorrect in general; $\sin(z)$ satisfies the linear homogeneous second order equation $f''(z)=-f(z)$ and vanishes
at many distinct points without being identically zero: 
$\sin(z)$ solves the two-point boundary value problem 
$f(z_1)=f(z_2)=0$ e.g. for $z_1=0$, $z_2=\pi$.
However, in the present case it is not required that $G_\pm(x;z)$ vanishes at two points in $D_0$ but that {\it both} components
of $\bm{f}(z)=(G_\pm(x;z),-G_\pm(x;-z))^T$ vanish at a single point $z_0\in D_0$. 
Because $0\in D_0$, the condition $G_\pm(x;0)=0$ is necessary and sufficient for $x-g^2$ to be an eigenvalue of $H_R$. 
Maciejewski {\it et al.} write: `` Numerical work seems to suggest that the condition at zero is somehow distinguished, \ldots''. The reason for this distinction is a simple mathematical fact explained above in great detail. 

Whereas isolated solutions of $G_\pm(x;z_0)=G_\pm(x;z_0^\ast)=0$ may exist for
some $z_0\in D_0$, it cannot happen for  $z_0\in i\R$, because then
$z_0=-z_0^\ast$ and 
$G_\pm(x;z_0)=0$ is equivalent to (\ref{cond-1a}, \ref{cond-1b}). This is the interesting case
from a numerical point of view, as it allows to overcome 
instabilities in the computation of $G_\pm(x;z)$ for large $x$.
It is remarkable that $G_\pm(x;z_0)$ has zeroes in $x$ 
at the correct values even when
$z_0\in i\R$ is outside $D_0$. The nonzero values of
$G_\pm(x;z_0)$ depend then on the order at which the defining series (\ref{phi12a})
 and (\ref{phi12b})
are truncated, but the position of $x^\pm_n$ with $G_\pm(x^\pm_n;z_0)=0$
converges to the correct value $E^\pm_n+g^2$ for the following reason: 
The functions $\p_1(z_0)$ and $\p_2(-z_0)$ are holomorphic
in $\Cc$ exactly at $x^\pm_n$, which entails a convergent
series expansion in $z$ at this point, even for $z\notin D_0$.  
   \begin{figure}[ht!]
    \centering
     \vspace*{5mm}
    \includegraphics[width=\hsize]{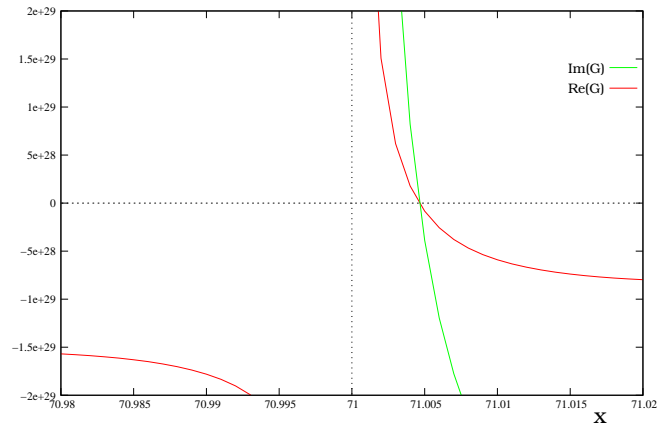}
    \vspace*{-0.3cm}
    \caption{Real and imaginary part of $G_+(x,5i)$ for $g=1$ and
$\D=0.7$ in the vicinity of $x=71$.}
    \vspace{-3mm}
    \label{Gi}
  \end{figure}
In Fig.~\ref{Gi} the real and imaginary part of $G_+(x;5i)$ is shown
for $x\approx 71$. $z_0$ is outside $D_0$ but the joint zero
of $\Re(G_+(x;5i))$ and $\Im(G_+(x;5i))$ allows
to determine the energy eigenvalue $E_{71}^+=70.00462935$ to high precision,
even though the series defining $G_+(x;5i)$ does not converge 
for $x\notin\{x_n^+\}$.  

We conclude that the  $z_0$ which may yield unphysical zeroes
of $G_\pm(x;z_0)$ are likely confined to the set $\R \cap D_0\setminus \{0\}$. 
This set is of measure zero
within $D_0$ \cite{note1}. 
In any case, its presence does not invalidate the results
of \cite{db}, which are based on the function $G_\pm(x)$ 
in Eq.~(\ref{sol}) and not
on the generalized function $G_\pm(x;z)$ for real $z\neq 0$, which is the subject
of criticism in \cite{mac}. The authors of \cite{mac} use the Wronskian
computed from $\p_1(z)$ and $\p_2(-z)$ together with their derivatives, 
which is equivalent
with the conditions (6) and (10) 
for the $\Zz_2$-invariant and the
generalized Rabi model, respectively.
The results of \cite{mac} neither correct nor extend the findings of \cite{db}
in these two cases, the only examples for which \cite{mac}
presents explicit calculations.    
Nevertheless, Maciejewski {\it et al.} have correctly pointed out a gap in 
the derivation of $G_\pm(x)$ \cite{supp}, giving me the opportunity to 
close it.


\begin{thebibliography}{99}

\bibitem{db} D.~Braak, Phys. Rev. Lett. {\bf 107}, 100401 (2011). 


\bibitem{mac} A.J.~Maciejewski, M.~Przybylska, and T.~Stachowiak, arXiv:1210.1130.

\bibitem{ince1} E.L.~Ince, {\it Ordinary Differential Equations}, Dover, N.Y.
(1956), p. 357.


\bibitem{supp} Online supplement to Ref.~\cite{db}.

\bibitem{mac2}  A.J.~Maciejewski, M.~Przybylska, and T.~Stachowiak, arXiv:1211.4639.

\bibitem{cod} E.A.~Coddington and R.~Carlson, {\it Linear Ordinary Differential Equations}, SIAM, Philadelphia (1997), p. 26.

\bibitem{ince2} see \cite{ince1}, Chap. IX.

\bibitem{note1} This conclusion is not affected by the possiblity that the two-point boundary value problem $G_\pm(z_0)=G_\pm(z_0^\ast)=0$ has  non-trivial solutions for isolated points
$z_0\notin \R \cup i\R$.


\end{thebibliography}
\end{document}